

\input epsf


\hfuzz .9cm
\rightline{UCLA/94/TEP/12}
\vskip .5in
\baselineskip=20pt
\centerline{\bf Extended BPH Renormalization of Cutoff Scalar Field Theories}
\vskip .5in
\centerline{\bf Gordon Chalmers}
\centerline{\it UCLA Physics Dept.}
\centerline{\it 405 Hilgard Ave., LA, CA 90024}
\vskip .4in
\centerline{Abstract}
\vskip .2in

{\narrower We show that general cutoff scalar field theories
in four dimensions are perturbatively renormalizable through
the use of diagrammatic techniques and an adapted BPH
renormalization method.  Weinberg's convergence theorem is
used to show that operators in the Lagrangian with dimension
greater than four, which are divided by powers of the cutoff,
produce perturbatively only local divergences in the two-,
three-, and four-point correlation functions.  We
also show that the renormalized Green's functions are the
same as in ordinary $\Phi^4$ theory up to corrections suppressed by
inverse powers of the cutoff.  These conclusions are
consistent with those of existing proofs based on the renormalization
group.}

\vfill
\vskip 3.0in
\noindent\hrule width 3.6in\hfil\break
{\it e-mail:chalmers@physics.ucla.edu}\hfil\break

\vfill
\break
\noindent {\bf I. Introduction}
\vskip .2in

The BPH (Bogoliubov-Parasiuk-Hepp) renormalization procedure$^{(1)}$ is
the oldest method by which
quantum field theories are renormalized.  Although this technique is
well known, it lacks a clear connection with the general
understanding given by the renormalization group, which in more
recent times has given a more physical understanding to renormalization.
Our purpose in this paper is to give a natural
extension of the BPH approach in order to show that general
cutoff field theories may be understood also by this more classic method.

Classic proofs of renormalization employing
the BPH method are based on the use of a recursive algorithm to
consistently remove all the divergences in the calculation of Greens
functions$.^{(2)(3)}$
Weinberg's convergence theorem is the most important ingredient$.^{(7)}$
It guarantees that at each order in perturbation theory
the new divergences are local in the external momenta, and therefore may
be removed perturbatively from all Greens functions by redefining the
couplings of (a finite number of) local operators in the
Lagrangian.  The types of theories amenable to this treatment are only
those containing interactions with couplings of non-negative dimension.

The main focus here is to deal with more general interactions, those
operators which have dimension greater than four
in the Lagrangian and which have couplings with inverse powers of
the cutoff $\Lambda$ (e.g. ${1\over \Lambda^2} \Phi^6,~{1\over
\Lambda^4} \Phi^5 \partial^2 \Phi$).  These operators, together with
the inverse cutoff factors, present problems to the classic proofs
of renormalization, as Weinberg's convergence theorem may
not be used directly to show that the divergences are local.  This will
be explained further below.  The
technique offered in this paper to circumvent this problem depends on
rearranging the Feynman integrals so that divergences of
graphs containing these higher dimensional vertices are shown to be local
(hence primitive).

Consider the example of a scalar Lagrangian containing the operator
${1\over \Lambda^2} \Phi^6$, with $\Lambda$ a momentum cutoff.  At
tree level such an interaction obviously vanishes as $\Lambda$ becomes
large.  Although a $\Phi^6$ term increases the degree of divergences
in Feynman diagrams, powers of the cutoff coming from the above vertex
suppress the final divergence in the loop integrations.  Naively, such
an interaction is power counting renormalizable, although one needs to be
more precise in justifying this.  The fault with naive power counting
arguments is that the ``mixing"
of the cutoff between the increased number of divergences (a result from
insertions of six-point vertices) with the vertex suppressions of
${1\over \Lambda^2}$ makes the use of Weinberg's theorem invalid.

Power counting may alternatively be justified in the framework of
the Wilson type renormalization group (RG)$.^{(4)}$  Indeed, these higher
dimensional operators in cutoff Lagrangians have been understood
for some time in terms of Wilson type RG flows.  The crux of this
method is that the scaling procedure naturally distinguishes between
operators of two types, relevant (also marginal) and irrelevant; and
the RG flow equations dictate that the only effect of the irrelevant
operators is to modify the physics with order $1\over \Lambda^2$
corrections.  Elegant proofs of renormalization have been put forth
using only the RG equations, with little mention of Feynman diagrams and
without using Weinberg's theorem at all$.^{(5)(6)}$  The general ``effective"
theory can then be shown to be perturbatively renormalizable,
essentially by dimensional analysis.
Contrasted with RG, the recursive BPH method is graphical in
nature and uses combinatorical arguments to handle the
potentially divergent Feynman diagrams.

The fact that the higher dimensional operators do give local
divergences is not as straightforward (as it is in RG) in terms
of diagrammatic perturbation theory, and will be elucidated in
what follows.  In this work we use the more traditional
techniques of BPH and Weinberg's convergence theorem to show
that general cutoff scalar field theories are perturbatively
renormalizable.

\vskip .3in
\noindent {\bf II. Renormalization Method}
\vskip .2in
\noindent{\it Standard~Theories}

The perturbative renormalization course must accomplish two tasks.
First, the Green's functions must be made well-defined, and second
the procedure must be outlined order by order as to carry out the
removal of divergences.  This problem, at least for standard power counting
renormalizable theories like $\Phi^4$ theory in four dimensions, was
solved many years ago.  The solution amounts to removing
all divergences associated with any set of internal loop momenta.
The Bogoliubov recursive formula is the mathematical statement:

$$
{\bar R}(G)=U(G)+ \sum_{ \{ \gamma_1,\ldots,\gamma_n \} \in P(G) } U
\biggl( {G \over \{ \gamma_1,\ldots,\gamma_n \} } \biggr) \prod^n_{i=1}
\biggl( -T_{ \gamma_i} {\bar R}(\gamma_i) \biggr), ~~\gamma_i \cap
\gamma_j = \emptyset \eqno{(1)}
$$
\vskip .2in

\noindent The set $\{ \gamma_1,\gamma_2,\ldots,\gamma_n \}$ ranges over
all partitions of 1PIR non-overlapping subgraphs of G.  $T_{\gamma}$
is a subtraction operator - Taylor expansion of degree $\delta(\gamma)$
in $\gamma$'s external momenta.  $U(\gamma)$ is defined to be the
value of the Feynman diagram, and ${\bar R}(\gamma)$ is the value
of $U(\gamma)$ with all of its subdivergences removed (in a recursive
manner).  We follow the notation of standard texts$.^{(2)(3)}$

In short, the expression (1) tells us to perform appropriate Taylor
expansions in the external momenta of all the subgraphs.  Weinberg's
theorem guarantees that the graph's divergent behavior is local and
contained in the finite order Taylor expansion.  The coefficients of the
expansion order by order in perturbation theory then are absorbed into the
definition of the coupling constants and correspond to the removal of
all the possible local divergences in the graphs that contribute to a
Green's function.
\vskip .3in
\noindent{\it General~Theories}
\vskip .3in
More general theories contain higher dimensional interactions; we
take the scalar Lagrangian (in Euclidean space) to be

$$
L= {1\over2}\partial\Phi\partial\Phi + {1\over2}m^2\Phi^2 -
{\lambda\over 4!}\Phi^4 + \sum_i {\lambda_i \over \Lambda^{2n-4}}
O_i^{(2n)} + L_{ct}, \eqno{(2)}
$$
\vskip .2in
\noindent where $O_i^{(2n)}$ are all operators $i$ with canonical
dimension $2n$ described below.  For simplicity we only consider
Lagrangians even under $\Phi \rightarrow -\Phi$.  This is a matter
of convenience, as there are fewer renormalization parameters
(by odd symmetry no 1- or 3-point counterterms), but our method
is easily extended.  The vertices $(\partial^{2n})(\Phi^{2m})$
with $n+m>2$ span all of the possible higher dimensional
Poincare-invariant operators in the Lagrangian.

The four dimensional theory is regulated by using a momentum
cutoff, and powers of this cutoff $\Lambda$ are used to
give the proper dimensions for the higher dimensional terms
in the theory.  The couplings for the higher dimensional operators
may also contain explicit ``soft dimensional" logarithms of
the cutoff (e.g., $\lambda_1 {\log ( {\Lambda\over m}) \over
\Lambda^2} \Phi^6$).  The following discussion is most
transparent if we explicitly exhibit the counterterm Lagrangian:

$$
L_{ct}={1\over2}(Z_{\Phi}-1)\partial\Phi\partial\Phi  +
{1\over2}(Z_m-1)m^2\Phi^2 - (Z_{\lambda}-1){\lambda\over 4!}\Phi^4 ,
\eqno{(3)}
$$
where $Z_{\Phi}$, $Z_m$, and $Z_{\lambda}$ are the counterterms adjusted
perturbatively to renormalize the 2- and 4-point functions.
\vskip .2in

The topology of a graph $G$ containing $v$ vertices of the type
${\Phi}^4$, $v_{2m}^{(2n)}$ vertices of the type
$(\partial^{2n})(\Phi^{2m})$ with $n+m>2$
(counted with derivatives and fields in any order), $I$ internal
lines, and $E$ external lines gives the topological relations:

$$
4v+\sum_{m+n>2} (2m)~ v_{2m}^{(2n)} = 2I+E, \eqno{(4)}
$$
$$
L=I-v+1-\sum_{m+n>2} v_{2m}^{(2n)} \quad \hbox{\rm $L$~$\equiv$~number~
of~loops} \eqno{(5)}
$$
\vskip .2in

\noindent We have separated the marginal operator $\Phi^4$ from
the irrelevant ones by counting $v$ and $v_{2m}^{(2n)}$ separately.  The
graph $G$ then has a superficial degree of divergence, $\delta(G)$,

$$
\delta(G)=4-E+\sum_{\rm G's~vertices} (2m+2n-4)
v_{2m}^{(2n)} \eqno{(6)}
$$
\vskip .2in

The same graph, though, comes with suppression factors of $1/
\Lambda$ raised to some power.  These factors enter through the
dimensions of the higher terms
in the cutoff Lagrangian
(e.g., ${1\over\displaystyle \Lambda^{2m+2n-4}} (\partial^{2n})
\Phi^{2m},~2m+2n>4$).  When the inverse powers of the cutoff are
considered within the power counting we see that the superficial
degree of divergence is lowered by an amount $\sum (2m+2n-4)
v^{(2n)}_{2m}$.  Effectively the power counting for the graph
G is modified to $\delta'(G)=4-E$, although {\bf $\delta$} is the
measure of the actual divergences from the loop integrations.  We
see, at least naively, that these theories are power counting
renormalizable, although we must first prove that the divergences
are local.  Considering the use of Weinberg's theorem, we must
carefully distinguish between $\delta$ and $\delta'$.  This is
essentially the crux of this paper.

\vskip .3in
\noindent{\it Weinberg's~Theorem}
\vskip .3in
Weinberg's convergence theorem establishes that if a graph $G$ has
$\delta(G)<0$, and all of its subgraphs $\Gamma$ have
$\delta(\Gamma)<0$, then $G$ is a well-behaved finite expression.
For primitively divergent graphs $G$, one has $\delta(G) \geq 0$, and
all of its subgraphs $\Gamma$ have $\delta(\Gamma)<0$.  Indirectly
we ascertain that the divergence in $G$ is a polynomial of degree
$\delta(G)$ in its external momenta.

Simply replacing $\delta$ with $\delta'$ in the use of Weinberg's
theorem is not allowed since the divergence from loop integrations
is really measured by $\delta$.  Take for example a graph $G$ with
$\delta'(G) \geq 0$ and all
of whose subgraphs $\Gamma$ satisfy
$\delta'(\Gamma)<0$, but with $\delta(\Gamma) \geq 0$ for some
$\Gamma$.  The usual argument of differentiating and integrating
the graph with respect to its external momenta in order
to show that the net divergence is local does not apply.  Even though
all subgraphs have $\delta' <0$, some of them may still possess
loop divergences ($\delta\geq0$).  Thus after differentiating $G$
with respect to its external momenta a number $\delta'(G) +1$ times
we are not guaranteed a finite integral.  Then by integrating
the differentiated graph with respect to the external momenta we
may not be sure that the divergence is polynomial.

In other words, $\delta$, rather than $\delta'$, power counting is
pertinent to the convergence theorem.  When all the vertices of a
graph correspond to relevant/marginal operators ($\Phi^4$ vertices
only in the case of a $Z_2$-symmetric 4-dimensional scalar theory),
the superficial degrees of divergence $\delta$ and $\delta'$ are
equal.  In this case the naive power counting is justified by the
use of the convergence theorem since the cutoff enters calculations only
through divergent integrals.  We only face a problem using
Weinberg's theorem with graphs containing the higher dimensional vertices,
when $\delta(G)$ and $\delta'(G)$ are not equal.
\vskip .3in
\noindent{\it Synopsis}
\vskip .3in
In order to clarify the formalities of the proof, given in
section III, a summary of the method is presented below.  In renormalizing
the theory using the BPH method (as opposed
to scaling arguments), the inverse cutoff factors from the vertices are
distinguished from the loop structure of a Feynman diagram, as:
\vskip .2in
$$
U(G)=
({1\over \Lambda^{\delta - \delta'} })
\cdot
\int ( \prod_{j_1}^{L} d^{d}l_{j} ) I(l_{1}, \ldots, l_{L};
p_1, \ldots, p_E )
$$
$$
\delta(G)-\delta'(G) = \sum_{\rm G's~vertices}
(2m+2n+4)v_{2m}^{(2n)}
\eqno{(7)}
$$
\vskip .2in
We then re-organize the loop structure and its $\delta'$ subtractions
into many integrals -- each one of which has at worst a primitive divergence
in view of Weinberg's theorem.  By keeping careful record of the
cutoff factors the divergences are shown to be local using $\delta'$.
Thus the process involves rearranging Feynman integrals, the use
of Weinberg's theorem to show that the divergences are local, and
then counting powers of the cutoff.  This procedure ultimately shows
that using Weinberg's theorem $\it{initially}$ with $\delta'$ counting is
justified.

We start by noting that in order to renormalize a divergent diagram
we have to perform all necessary subtractions to the original graph G to
render its sub-integrations finite.  This is done with $\delta$ power
counting and includes all the counterterms pertaining to
the higher dimensional vertices.  Weinberg's theorem
tells us that the net divergence in the loop integrations of the
graph G plus its lower order counterterms is a polynomial in G's
external momenta, and that in general a truly primitive graph with
a superficial degree of divergence $\delta$ will then require
counterterms from all operators of dimension $\delta$ or less.

The greater than four dimensional counterterms are by dimensional
analysis, however, divided by the appropriate power of $\Lambda$
(i.e. $f(\ln \Lambda/m ) \over\displaystyle
\Lambda^{2m+2n-4}$ for a $\lambda_i (\partial^{2n})(\Phi^{2m})$ vertex) -
they are also suppressed.  These counterterms
are added to the bare Lagrangian (or likewise perturbatively to the
graphs), but then subtracted back out.  The subtracted terms
generate sets of new graphs with fewer loops and simpler topological
structure, which may be analyzed through the same procedure.  As
we will see, only the 2- and 4-point counterterms need to be
explicitly added perturbatively to the Greens functions.

A very simple example is illustrated in figure 1.  Figure 1(a) shows a
divergent one loop contribution to
the 6-point function which, for the sake of argument, could be
embedded into some larger graph.  By adding and subtracting the
divergent piece, illustrated graphically in figure 1(b), we re-organize
the original diagram into a finite integral plus a tree-level
interaction.  The tree-level interaction (suppressed polynomial
subtraction) was not only added to
the original graph, but added and subtracted.  Weinberg's
theorem forces the expression in figure 1(a) to split into two
pieces:

$$
{1\over \Lambda^2}\Big[ \delta(G)=0~\hbox{\rm dimensional~(finite)
{}~function} \Big] +
{1\over \Lambda^2}\Big[
\hbox{\rm polynomial~divergence~of~degree~} \delta(G)=0 \Big]
\eqno{(8)}
$$
\vskip .2in

\noindent Here the cutoff factor from the ${1\over \Lambda^2}
\Phi^6$ vertex has been separated from the Feynman integrals.

We continue this procedure until the general graph $G$ and its
lower order counterterms according to $\delta'=4-E$ power
counting are transformed into sets of graphs $G_j$ plus counterterms
according to $\delta$ power counting.  Each set now has all
subdivergences in loop integrations removed; the net divergent
behavior of G is local in view of the convergence theorem.  In the
notation used in this paper, we have that ${\bar R}'(G)=\sum
{\bar R}(G_j)$; each ${\bar R}(G_j)$ contains no divergent
sub-integrations and has only a net primitive divergence, so
${\bar R}'(G)$ (the graph with all of its lower order counterterms
according to $\delta'$ counting) also has at worst a local divergence.

This inductive method relies upon adding and subtracting many
counterterms to the original graph.  Each of which vanish at least
as fast as $1/ \Lambda^2$ as $\Lambda\rightarrow\infty$.  This
procedure re-organizes the Feynman integrals so that the divergent
structure of the perturbative Greens functions may be dissected
through using Weinberg's theorem.  In this manner the usual
renormalization tools may be used to justify $\delta'=4-E$ power
counting.  By keeping track of the cutoff factors,
the $O({1\over \Lambda^2})$ bounds on the renormalized Green's functions
are also found.  (By further bookkeeping on the number of loops in the
perturbative expansion, logarithmic corrections to
the bounds may in principle be found.  We shall not do so here.)

The conclusions of this analysis are summarized below:
\vskip .25in

1.  A general diagram G which has all of its subdivergences removed
according to $\delta'$ power counting has at worst a divergence which
is a polynomial in the external momenta of degree $\delta'(G)$.  This
is analagous to Weinberg's theorem but with the modified power counting.

2.  A renormalized graph containing at least one of the higher-
dimensional vertices is at most proportional to $E^2/ \Lambda^2$
(times $\log$s) and vanishes as $\Lambda \rightarrow \infty$.  This is
with the physical scale E $(p_i^2 < \Lambda^2)$ of the Green's
functions fixed below $\Lambda$.

\vskip .6in
\noindent {\bf III. Proof of Renormalization}
\vskip .2in

The proof will be broken up into three steps and will follow the BPH
method in organizing the renormalization.  In section (a) we deal with the
simplest graphical structures - namely disjoint renormalization parts
which do not contain subdivergences.  We re-organize the subtractions
as described previously and show that only the two- and four-point
Greens functions need renormalization.  This part most clearly
demonstrates the adding and subtracting procedure, allowing us
to effectively use Weinberg's theorem but with $\delta'$ power
counting.  In section (b) the subtractions are organized that render graphs
finite, but only for those which do not contain nested subdivergences
(more explicitly, only graphs whose one particle irreducible, 1PIR,
subgraphs do not themselves
contain further subdivergent integrations).  The point here is to
build the recursion formula and to illustrate once again how the
adding and subtracting process is used to show that the divergences
are local.  In section (c) we deal with general Feynman
diagrams.  In this section the ${\bar R}'$ operation is utilized on
the subgraphs to successively remove the subdivergences according to
$\delta'$.  The rearrangement is used to show how ${\bar R}'(G)$
of a general graph G breaks into a specific form, a sum of
primitively divergent integrals, $\sum{\bar R}(G_i)$, thus
justifying that the only counterterms necessary are those by
$\delta'$ counting.  In this manner, the Bogoliubov recursion
formula will be derived but with $\delta'$ subtractions.

\vskip .4in

\noindent $\underline{(a)~Disjoint~Subgraphs}$
\vskip .2in

To this extent, consider first the subgraph $\Gamma$ below in
figure 2(a).  $\Gamma$ is made up of two disjoint parts $\gamma_1$
and $\gamma_2$ such that $\Gamma=\gamma_1 \cup \gamma_2$ and
$\gamma_1 \cap \gamma_2 = \emptyset$.  Both $\gamma_1$ and $\gamma_2$
themselves have no subdivergences, hence they have at worst
primitive divergences of degrees $\delta(\gamma_1)$ and
$\delta(\gamma_2)$.  Following Dyson's prescription the divergence
from the loop integrations in $\Gamma$ is removed by replacing:

$$
U(\Gamma) \rightarrow (1-T_{\gamma_1})(1-T_{\gamma_2})U(\Gamma)
\eqno{(9)}
$$
\vskip .2in

\noindent The operation $T_{\gamma}$ is an order $\delta(\gamma_1)$
Taylor expansion in the external momenta of $\gamma_1$, and
corresponds to including the appropriate counterterms in the bare
Lagrangian.

More general subgraphs $\Gamma$ have many disjoint components, and may be
broken into several disconnected 1PIR components $\gamma_i$ so that
$\Gamma=\cup^n_{i=1} \gamma_i,~\gamma_i\cap\gamma_j=\emptyset$.  Assume
for now that all of the $\gamma_i$ have no subdivergences (i.e.,
$U(\gamma_i)={\bar R}(\gamma_i)$), so that they
all have at worst primitive divergences.  All divergences coming from
the loop integrations are eliminated by replacing (as in figure 2b):

$$
U(\Gamma)\rightarrow \prod^n_{i=1} (1-T_{\gamma_i})U(\Gamma)\equiv
R(\Gamma) \eqno{(10)}
$$
\vskip .2in

When the subgraph contains higher dimensional vertices, this
expression is excessively oversubtracted.  $T_{\gamma}$ is a Taylor
expansion of degree
$\delta(\gamma)$ not $\delta'(\gamma)$, with
some $\delta(\gamma_i)>\delta'(\gamma_i)$.  This
means that all counterterms, including the ones which would
renormalize the irrelevant operators, have been added to make
{\it all} sub-integrations finite.  This is clearly unnecessary by
dimensional analysis, in view of the overall factor $1/
\Lambda^{2m+2n-4}$ in front of these higher dimensional vertices
$(\partial^{2n})\Phi^{2m}$ ($m+n>2$).  Some divergences are ``eaten up"
by the vertex suppressions.  To re-iterate this point,
the most general counterterm to these higher dimensional
operators by dimensional analysis
has the form $f(\ln{\Lambda\over m})/ \Lambda^{2m+2n-4}$,
and vanishes as $\Lambda \rightarrow \infty$.  The
only subtractions we want to make are according to $\delta'$ power
counting.  With this in mind, the necessary change in the above
prescription is then to use $\delta'$:

$$
U(\Gamma)\rightarrow \prod^n_{i=1} (1-{T'}_{\gamma_i})U(\Gamma)
\equiv R'(\Gamma) \eqno{(11)}
$$
\vskip .2in

\noindent This primed expression means that in subtracting the
subgraph $\Gamma$ only the bare four-point and two-point parameters
need adjustment (since $\delta'=4-E$), or likewise only the
counterterms to $Z_m$, $Z_{\lambda}$, and $Z_{\Phi}$ need to be
adjusted.  Now we show that this expression truly yields
a finite result.

Following the discussion in the introduction let's add and subtract
the unnecessary renormalizations and rewrite $R'(\Gamma)$ in (11) as:

$$\eqalign{
R'(\Gamma) & = \prod^n_{i=1} (1-{T'}_{\gamma_i})U(\Gamma)=\prod^n_{i=1}
\biggl\{ 1-T_{\gamma_i}+T_{\gamma_i}-{T'}_{\gamma_i} \biggr\}
U(\Gamma)
\cr &
=\sum_{\rm partitions} \prod_{\gamma_j \in A} (1-T_{\gamma_j})
\prod_{\beta_k \in B} (T_{\beta_k}-T_{\beta_k}')U(\Gamma) }
\eqno{(12)}
$$
\vskip .2in

\noindent The sum extends over the ways in which the IPIR parts of
$\Gamma$ may be
partitioned into two sets $A=\cup \gamma_j$ and $B=\cup \beta_k$ such
that $\Gamma=A \cup B$.  More explicitly (12) may be expanded,

$$
R'(\Gamma)=\prod^n_{i=1} (1-T_{\gamma_i})U(\Gamma) + \sum_{j=1}^n
\bigl\{ \prod_{i\neq j} (1-T_{\gamma_i})
\bigr\} \biggl\{ (T_{\gamma_j}-{T'}_{\gamma_j})U(\Gamma) \biggr\}+\ldots
\eqno{(13)}
$$
\vskip .2in

The $(1-T_{\gamma})$ subtraction eliminates the loop divergence
coming from the integration in $U(\gamma)$, and individually the
operation $(T_{\beta}-{T'}_{\beta})U(\Gamma)$ in (13) replaces the 1PIR
renormalization part $\beta$ in $\Gamma$ with a polynomial in its
external momenta.  The $(T_{\beta}-T_{\beta}')$ operation when
acting on the primitively divergent 1PIR graph $U(\beta)$ in fact always
results in a polynomial with terms divided by powers of at least $\Lambda^2$.
\vskip .5in
\noindent{\it $T-T'$~Operation}
\vskip .2in
The fact that $T-T'$ is proportional to at least $1/ \Lambda^2$ is
found by carefully separating the inverse cutoff factors that
arise from the vertices in $\beta$ from its loop momentum
structure.  The vertices in $\beta$ contribute the cutoff factor:

$$
f_{\beta}(ln{\Lambda\over m}) {1\over \Lambda^{\delta(\beta)-
\delta'(\beta)}} \eqno{(14)}
$$
$$
\delta(\beta) - \delta'(\beta) =
\sum_{\rm \beta 's~vertices} (2m-2n-4)v_{2m}^{(2n)}
$$
\vskip .2in

\noindent to the value of the renormalization part $\beta$ (where
$f_{\beta}$ is a product of possible logarithmic factors in the
tree-level vertices), and the operation $(T_{\beta}-
T'_{\beta})U(\Gamma)$ on the primitively divergent
loop integral above is a polynomial of the form:

$$
g_1(ln{\Lambda\over m}) p^{\delta(\beta)} + g_2(ln{\Lambda\over m})
\Lambda^2 p^{\delta(\beta)-2}+
\ldots+g_k(ln{\Lambda\over m})
\Lambda^{\delta(\beta)-\delta'(\beta)-2} p^{\delta'(\beta)+2}
\eqno{(15)}
$$
\vskip .2in

\noindent Recall that $T$ and $T'$ are  Taylor expansions of order
$\delta$ and $\delta'$ respectively, and the coefficients $g_k$ are
logarithmic factors of the cutoff which arise from the divergence
of the 1PIR renormalization part $\beta$.

The value of $(T_{\beta} -T'_{\beta})U(\Gamma)$ then has the form
(combining the vertex factors and the above polynomial subtractions),

$$
{f_{\beta}(ln{\Lambda\over m})\over \Lambda^{\delta(\beta)-
\delta'(\beta)}} \Biggl( g_1(ln{\Lambda\over m})
p^{\delta(\beta)} + g_2(ln{\Lambda\over m})
\Lambda^2 p^{\delta(\beta)-2}+\ldots+
g_k(ln{\Lambda\over m}) \Lambda^{\delta(\beta)-\delta'(\beta)-2}
p^{\delta'(\beta)+2} \Biggr) U({\Gamma\over \beta}) \eqno{(16)}
$$
$$
\equiv \sum_{j=1} U'_j (\beta)U({\Gamma\over \beta}) \eqno{(17)}
$$
\vskip .2in

The form in (17) is written in a manner to show how the
$(T_{\beta}-T'_{\beta})$ operation effectively reduces the
renormalization part $\beta$ into several tree level interactions,
$\sum U'_j (\beta)$, given by the polynomial above in (16).  The
sum of terms in (16)  (the polynomial times the vertex factors)
contain only terms monomial in $\beta$'s external momenta divided
by powers of the cutoff, which is similar to replacing the subgraph
$\beta$ under this operation with a sum of higher dimensional
suppressed vertices.  In general we have products of the $T-T'$
operations acting on a set $\{ \beta_k \}$, as in (12); then each
of the disjoint $\beta_k$ is replaced with several tree-level
vertices - all of them divided
by at least $\Lambda^2$.  Lastly it is important to note that $T-T'$
acting on {\it any} primitively divergent set of graphs ${\bar R}(G)$
is a polynomial whose terms are divided by at least $\Lambda^2$.
\vskip .3in
\noindent{\it $\delta '$~Subtractions}
\vskip .2in
The effect of changing $(1-T'_{\beta})
\rightarrow (1-T_{\beta}) +
(T_{\beta}-T'_{\beta})$ allows the divergent behavior to be
understood in view of Weinberg's theorem.  Take one of the terms
in the expansion given in (12), where we denote one of the partitions
of $\Gamma$ into two sets $\Gamma_1$ and $\Gamma_2$ by the index $p$:

$$
R'(\Gamma_p)=\prod_j (1-T_{\gamma_j})\left\{ \prod_k (T_{\beta_k}-
T'_{\beta_k}) U(\Gamma) \right\} \eqno{(18)}
$$
where,
$$
\Gamma_{p}=\{ \gamma_j \} \cup \{ \beta_k \} = \Gamma_1 \cup \Gamma_2
$$
\vskip .2in

\noindent Since all of the elements $\gamma_j$ and $\beta_k$ are
disjoint from one another, all the subtractions commute and we are
left with two functions multiplying each other.  We write (18) as the
product below in order to illustrate this point:

$$
R'(\Gamma_p) = \Biggl( \prod_j (1-T_{\gamma_j})U(\Gamma_1) \Biggr)
\cdot U'(\Gamma_2) \eqno{(19)}
$$
where,
$$\eqalign{
U'(\Gamma_2) & = \prod_k(T_{\beta_k} -T'_{\beta_k})U(\Gamma_2)
\cr &
=\prod_k \Biggl( \sum_{i_k =1} U'_{i_k} (\beta_k) \Biggr) }
\eqno{(20)}
$$
\vskip .2in

The original set of disjoint renormalization parts of $\Gamma$ has
been partitioned into two new sets $\Gamma_1=\{\gamma_k\}$ and
$\Gamma_2=\{ \beta_j\}$.  All of the elements $\beta$ in $\{ \beta_k\}$
have been replaced with tree level interactions, $\sum_{i_k} U'_{i_k}
(\beta_k)$, in view of the previous discussion, and the remaining
parts $\Gamma_1=\{\gamma_k\}$ have their divergences in the loop
integrations completely removed according to Dyson's prescription
(i.e., subtractions according to $\delta$ power counting).

Note that $U'(\Gamma_2)$, written out in (20), is a sum of many terms
found by factoring the product of the polynomials:

$$
U'(\Gamma_2)=\prod_k \Biggl( \sum_{i_k =1} U'_{i_k} (\beta_k)
\Biggr) = \sum_{all~sets~\{ i_1,i_2,\ldots\} } \prod_k U'_{i_k}
(\beta_k) \eqno{(21)}
$$
\vskip .2in

\noindent This means that out of
every polynomial $(T_{\beta_k}-T'_{\beta_k})U(\beta_k)$ for all $\beta_k \in
\Gamma_2$ we take one term $i_k$, then we sum over all the
distinct ways of doing this.

The expression (18) is thus a finite
function ($\delta(\Gamma)-\delta'(\Gamma)=0$) or a finite function divided
by powers of the cutoff since for each $\Gamma_p$:

$$
R'(\Gamma_p)=U'(\Gamma_2) \cdot { f_{\Gamma_1}
\over \Lambda^{\delta(\Gamma_1)-\delta'(\Gamma_1)} }
\cdot \left[ \delta(\Gamma_1) \hbox{ \rm ~dimensional~finite~function}
\right] \eqno{(22)}
$$
\vskip .2in

The factors $\Lambda^{\delta(\Gamma_1)-\delta'(\Gamma_1)}$ and
$f_{\Gamma_1} (ln{\Gamma\over m})$ come from the vertices (since
the couplings may have explicit logarithms) in $\Gamma_1$
and even further suppress the polynomial terms in $U'(\Gamma_2)$.
The only case in which the expansion in (12) does not lead to terms
(18) divided by at least $\Lambda^2$ is when there are no
high-dimensional vertices in $\Gamma$, $\delta(\Gamma)=\delta'(\Gamma)$.
This is when the usual BPH renormalization procedure is regained.

Consider the subgraph in Figure 3(a).  The subtraction according to
$\delta'$ in this example splits into two pieces (figure 3b):

$$
(1-T'_{\gamma_1})(1-T'_{\gamma_2}) U(\Gamma) ={1\over \Lambda^2} \Biggl(
\delta(\Gamma)=0 \hbox{ \rm ~dim.~finite~function} \Biggr)
$$
$$
+( {\ln ({\Lambda \over m})
\over \Lambda^2}) \cdot
\Biggl( \delta(\gamma_1)=0 \hbox{ \rm ~dim.~finite~function} \Biggr)
\eqno{(23)}
$$
\vskip .2in

\noindent In the figure, the dashed and solid boxes represent the
$T-T'$ and $T$ operations on the two renormalization parts
$\gamma_1$ and $\gamma_2$.

Summing over all the ways in which the original $\Gamma=\{ \gamma_j \}$
may be split into two sets $\Gamma_1 = \{ \gamma_k \}$ and
$\Gamma_2 = \{ \beta_j \}$ generates the terms in (12), each one
of which has the form above in (22).  If $\Gamma$ contains a
high-dimensional vertex, than each element in the expansion is
divided by at least $\Lambda^2$.

To summarize so far, we see that the prescription according to
$\delta'$ counting (a modified Dyson's prescription),

$$
U(\Gamma) \rightarrow \prod_{i=1}^n (1-T'_{\gamma_i})U(\Gamma)
\eqno{(24)}
$$
\vskip .2in

\noindent leads to finite expressions (as
$\Lambda \rightarrow \infty$).  In the case that the subgraph
$\Gamma$ contains one of the higher dimensional vertices these
results are divided by powers of the cutoff.  Note also that if a
subgraph $\gamma$ satisfies $\delta(\gamma)=\delta'(\gamma)$,
which means no irrelevant operators in $\gamma$, then no
substitutions take place ($(T_{\gamma}-T'_{\gamma})U(\Gamma) = 0$).  We
regain the conventional BPH prescription using $\delta$ power
counting on graphs containing no higher dimensional vertices.

\vfill\break
\noindent $\underline{(b)~Non-Nested~Divergences}$
\vskip .3in

Now we consider Feynman diagrams which contain many subgraphs but
no nested subdivergences.  According to $\delta'$ power counting the
overall divergence from a particular subgraph $\Gamma$ in G is removed
by subtracting it from the graph as follows:

$$
\prod_{\gamma\in\Gamma} (1-T'_{\gamma}) U(G)= U(G) + \sum_{\beta\in
P(\Gamma)} U\Bigl( {G\over \{\beta_1,\beta_2,\ldots,\beta_n\} }
\Bigr) \prod_k (-T'_{\beta_k}) U(\beta_k) \eqno{(25)}
$$
\vskip .2in

\noindent where $\Gamma=\cup \gamma_i$ and $P(\Gamma)$ extend over the
partitions of $\Gamma$ into sets $\{ \beta_k \}$, not counting the
null set.  Next, the $\delta'$ divergences from all of $G$'s
subgraphs are removed.  The subtractions from all of $G$'s subgraphs
are summed with the condition that we count the equivalent ones only
once.  This amounts to summing only over the partitions of G (this
set generates all the partitions of every subgraph only once):

$$
{\bar R}'(G)= U(G) +
\sum_{\beta\in P(G)} U\Bigl(
{G\over \{\beta_1,\beta_2,\ldots,\beta_n\} } \Bigr) \prod_k (-T'_{\beta_k})
U(\beta_k) \eqno{(26)}
$$
\vskip .2in

For the moment assume that the 1PIR renormalization parts $\beta_k$
do not themselves contain loop subdivergences.  Equation (26) is
then the expression for the graph
G with all of its subdivergences removed according to $\delta'$
counting.  The usual BPH arguments use Weinberg's theorem to deduce
that the overall divergence of (26) is a local polynomial.  However,
 certain sub-integrals in the above may still be divergent -- just
divided by powers of the cutoff since we are counting with $\delta'$.
Thus we are not yet in a position to say that ${\bar R}'(G)$ has a
local divergence.  By re-organizing the above form we intend to show
that it truly does have a local divergence in view of the convergence
theorem.

Split the subtraction in equation (26) and we have:

$$\eqalign{
{\bar R}'(G) & = U(G) +
\sum_{\beta\in P(G)} U\Bigl(
{G\over \{\beta_1,\beta_2,\ldots,\beta_n\} } \Bigr) \prod_k (T_{\beta_k}-
T'_{\beta_k} - T_{\beta_k}) U(\beta_k)
\cr &
=\sum_{\Gamma_j} U'(\Gamma_j) \Biggl\{ U({G\over \Gamma_j}) +
\sum_{ \{ \gamma_1,\ldots,\gamma_n \} \in P( {G\over \Gamma_j})}
U({G\over \{ \Gamma_j,\gamma_1,\ldots,\gamma_n \} })
\prod_{i=1}^a (-T_{\gamma_i} U(\gamma_i)) \Biggr\} }
\eqno{(27)}
$$

$$
with~~U'(\Gamma_j) \equiv \prod_{i=1}^n (T_{\beta_i}-T'_{\beta_i})
U(\beta_i), \quad \Gamma_j = \cup_{i=1}^n \beta_i
$$
\vskip .2in

\noindent (The sum over $\Gamma_j$ is over all
partitions of G ($\Gamma_1=\emptyset,\Gamma_2=\{\gamma_1\},
\{\gamma_1,\gamma_2\},
\ldots,\Gamma_n=\{G\}$) including G, and the next sum extends over
 all the partitions of $G\over \Gamma_j$ not including $G\over \Gamma_j$).

The factorization expressed in (27) is rather direct and follows
from the structure of G - that all of its 1PIR components $\gamma_i$
in every subgraph do not themselves contain further subdivergences
(according to $\delta$).  $U(\gamma_i)$ is at most primitively
divergent for all $\gamma_i$.  In more general cases we need to
first remove the subdivergences in $\gamma_i$ recursively, namely
with the ${\bar R}'(\gamma_i)$ operation.

Equation (27) expresses the re-organization of a Feynman diagram
G together with its $\delta'=4-E$ counterterms into a set of new
diagrams with subtractions given by $\delta$ counting.  The sum
over the partitions $P({G\over \Gamma_j})$ together with the
subtraction $T$ renders all subintegrations in $G\over \Gamma_j$
completely finite.  Individually each set denoted by
$G\over \Gamma_j$ contains only true primitive momentum divergences
in view of Weinberg's theorem.

The grouping of the subtractions in equation (27) should be
contrasted with that of (26).  Previously the sub-integrations
in G may be divergent, just divided by hard powers of the cutoff.  But
by adding and subtracting the extra terms we arrive at Feynman
integrals which have completely convergent subgraphs, but also
divided by powers of the cutoff.  Weinberg's theorem may now be
used to justify that the overall divergence in ${\bar R}'(G)$ is
local since every subgraph $\gamma$ of $U'(\Gamma_k)U({G\over \Gamma_k})$
together with its corresponding counterterms  has effectively
$\delta(\gamma) < 0$.

Take as an example a graph with three 1PIR parts
($\gamma_1 ,\gamma_2 ,\gamma_3$), where each may be at most
primitively divergent.  In one term of
the above expansion the $T-T'$ operates on the part $\gamma_3$; the
$U'(\{\gamma_3\})$ is thus a polynomial in
$\gamma_{3}$'s external momenta as discussed in the first part of the
paper.  The graph $U({G\over \{\gamma_3\}})U'(\{\gamma_3\})$
contains three divergences, and they have all been
subtracted out by the operation:

$$
\sum_{\Gamma\in P( {G\over \{\gamma_3\}})} U({G\over
\{ \Gamma,\gamma_3\}}) \prod_{\beta\in\Gamma} (-T_{\beta})
U(\beta)
$$
$$
=(-T_{\gamma_1}-T_{\gamma_2} + T_{\gamma_1}T_{\gamma_2})
U'(\gamma_3)U(\{ \gamma_1,\gamma_2\} )
\eqno{(28)}
$$
\vskip .2in

\noindent By Weinberg's theorem the net divergence in the four
diagrams, including $U'(\gamma_3) U({G\over \gamma_3})$, must be a
polynomial in $G$'s momenta.  The
finite part of the integral is divided by whatever factors of the
cutoff are present, those in the piece $U'(\gamma_3)$ and the vertices in
$U({G\over\gamma_3})$, and is at least $\Lambda^2$ since $T-T'$
produces suppressions proportional to $1/ \Lambda^2$.

In more detail let's inspect the re-organized expression in
equation (27), which we write in a more revealing notation:

$$
{\bar R}'(G)= \sum_{G_k}\Biggl( U(G_k) + \sum_{ \{\gamma_1,
\ldots,\gamma_n\} \in P({G_k})} U\Bigl( {G\over \{\gamma_1,
\ldots,\gamma_n\} } \Bigr) \prod_{i=1}^n (-T_{\gamma_i}
U(\gamma_i) )\Biggr), \eqno{(29)}
$$
where
$$
U(G_k)=\Bigl( \prod_{i=1}^a (T_{\beta_i}-T'_{\beta_i})
\Bigr) U({G\over \{\beta_1,\ldots,\beta_n\} }),~~and~~\{\beta_1,
\ldots,\beta_n\}\in P(G)
$$
\vskip .2in

Each $G_i$ above is a graph containing many 'new' vertex insertions
 and has the same topological structure as G except for being at
least one loop less.  The $T-T'$ operation has replaced particular
loops in the original graph G with polynomials in the loop's
external momenta.  Each polynomial leads to a number of terms
divided by at least $\Lambda^2$, as described previously in
section III(a).  These new graphs should then further be denoted
by all possible combinations of the new vertices (the terms in the
suppressed polynomial) inserted from the $T-T'$ acting on
different renormalization parts.  Considering Weinberg's theorem,
we may write (29) as:

$$
{\bar R}'(G)={\bar R}(G)+\sum {\bar R}(G_i) \eqno{(30)}
$$
\vskip .2in

\noindent a sum of primitively divergent graphs.

The convergence theorem tells us indirectly that each ${\bar R}
(G_i)$ has at worst a divergence which is polynomial.  The simplest
piece in (29) is found by taking $\Gamma_k = \{\emptyset\}$:

$$
{\bar R}(G)= U(G) + \sum_{P(G_k)} U\Bigl( {G\over \{\gamma_1,
\ldots,\gamma_n\} } \Bigr) \prod_{i=1}^n \Bigl(-T_{\gamma_i}
U(\gamma_i) \Bigr) \eqno{(31)}
$$
$$
={f(ln{\Lambda\over m})\over \Lambda^{\delta(G)-\delta'(G)}}
\Bigl( \hbox{ \rm
divergent~polynomial~of~degree~$\delta$(G)~+~finite~function~of~
dim~$\delta$(G)} \Bigr)
$$
\vskip .2in

\noindent The overall cutoff factor coming from the higher vertices
in G has been extracted from the original structure of the graph.
The form in the brackets is given by the locality of the divergence,
since all necessary counterterms to render $G$'s subintegrations
finite have been included.  After subtracting out the local
divergence in ${\bar R}(G)$ (of overall degree $\delta'$), this
entire quantity must be proportional to the hard vertex cutoff factors
in G, $1\over \Lambda^{\delta(G)-\delta'(G)}$, times the remaining
terms in the polynomial and the finite function.  This is at most
$1/\Lambda^2$ since $\delta(G)-\delta'(G)\geq 2$ for a graph G
containing at least one higher dimensional vertex.  The function
$f(\ln {\Lambda\over m})$ comes from the possible logarithmic
factors in the vertices and gives only a small correction to the
hard powers of the suppression.

The remaining terms in (29) are more complicated owing to the
$T-T'$ operation.  For example, suppose $(T-T')$ acting on a
subgraph $\gamma$, $(T-T')U(\gamma)$, leads to three terms:
${m^4\over\Lambda^4} \ln{\Lambda\over m},~{\Lambda^2 m^2
\over\Lambda^4} \ln{\Lambda\over m},~{\rm and}~{p^4\over \Lambda^4}
\ln{\Lambda\over m}$.  Each one may be thought of as a vertex in
the diagram $G_j$.  The last has an internal momentum flowing through
and leads to a higher degree of divergence in the overall integration
in $G_j$ than the first two terms.  The second term however gives
a result an order in $\Lambda^2$ higher than if we had
inserted ${m^4\over\Lambda^4}ln{\Lambda\over m}$ in its place.

As a result, the structure of ${\bar R}(G_j)$ will break up into
primitive divergences of varying degrees.  The matter is complicated
when we consider that there are many of these polynomials in any
reduced diagram, all from the many $T-T'$ operations
acting on the disjoint renormalization parts (in the $U'(\Gamma_k)$
found in equation (29) for example).  This is illustrated in figure 4
where the dashed boxes around the subgraphs represent the $T-T'$
operation.  Notationally each $G_j$ and its counterterms actually represent
many diagrams as mentioned before, labeled by choosing one term out
of each polynomial from the $T-T'$.

The product of the disjoint $T-T'$ operations,
$\prod^a_{i=1}(T_{\beta_i}-T'_{\beta_i})U(\beta_i)$,
found in (29) factors into a sum, as in the previous section of
the paper (eqns. (20)-(21)).  Define $U'_{n_i}(\beta_i)$ to be
the $n_i^{th}$ term in the polynomial $(T_{\beta_i}-T'_{\beta_i})
U(\beta_i)$.  Then,

$$
\prod^a_{i=1} (T_{\beta_i}-T'_{\beta_i})U(\beta_i)=\prod^n_{i=1}
\Biggl( \sum^{N_i}_{i=1} U'_{n_i}(\beta_i) \Biggr)
$$
$$
=\sum_{{\rm all~sets}~\{n_1,\ldots,n_a\}} \prod^a_{i=1} U'_{n_i} (\beta_i)
\eqno{(32)}
$$
\vskip .2in

\noindent Summing over the sets $\{n_1,\ldots,n_a\}$ takes us over
all combinations of making a product out of '$a$' terms, one term $n_i$
from each of the polynomials.  The number N $(N=n_1 n_2 \cdots n_a)$
of distinct combinations is in general quite large, and depends on
how many terms $n_i$ there are in each of the $(T_{\beta_i}-
T'_{\beta_i})$.

The major difference here from section III(a) is that the terms
$\prod^a_{i=1} U'_{n_i} (\beta_i)$ are in fact embedded in the
graph $G_k$ in (29), and hence contribute to the Feynman
integrals.  Recall that the form of $U'_{n_i} (\beta_i)$ depends
on the renormalization part $\beta_i$, being one monomial out of
the following,

$$
{f_{\beta_i}(ln{\Lambda\over m})\over
\Lambda^{\delta(\beta_i)-\delta'(\beta_i)}}
\Biggl( g_1(ln{\Lambda\over m}) p^{\delta(\beta_i)}
+ g_2(ln{\Lambda\over m}) \Lambda^2 p^{\delta(\beta_i)-2}+
\ldots+g_k(ln{\Lambda\over m})
\Lambda^{\delta(\beta_i)-\delta'(\beta_i)-2}
p^{\delta'(\beta_i)+2} \Biggr)  \eqno{(33)}
$$
\vskip .2in

\noindent All the pieces in (33) are divided by at least $\Lambda^2$,
but carry different dimensions of momentum, mass, and the cutoff.
'$P$' refers generally to the momentum flowing through the vertex,
either internal loop momentum or off an external leg.  Each of the
distinct combinations in (32) thus gives a different internal loop
momentum structure and contributes differently to the overall divergence.

Explicitly, the general ${\bar R}'(G_k)$ must break into $N=n_1 n_2
\cdots n_a$ terms as follows:

$$\eqalign{
{\bar R}'(G_k) & = {h_1(ln{\Lambda\over m})\over
\Lambda^{\sum (2m+2n-4)v^{(2n)}_{2m}}}
\Biggl(\hbox{\rm divergent~polynomial~of~degree~$\delta(G)$~+
{}~finite~function~of~
dim.~$\delta(G)$} \Biggr)
\cr &
+{h_2(ln{\Lambda\over m})\over
\Lambda^{\sum (2m+2n-4)v^{(2n)}_{2m}}}
 \Lambda^2 \Biggl(\hbox{\rm divergent~polynomial~of~
degree~$\delta(G)-2$~+~finite~
function~of~
dim.~$\delta(G)-2$} \Biggr)
\cr &
+{h_3(ln{\Lambda\over m})\over
\Lambda^{\sum (2m+2n-4)v^{(2n)}_{2m}}}
 m^2 \Biggl(\hbox{\rm divergent~polynomial~of~
degree~$\delta(G)-2$~+~
finite~function~of~
dim.~$\delta(G)-2$} \Biggr)
\cr &
+\ldots+{h_n(ln{\Lambda\over m})\over
\Lambda^{\delta(G)-
\delta'(G)}}\cdot \prod^a_{i=1} \Lambda^{\delta(\beta_i)-
\delta'(\beta_i)-2} \Biggl( \hbox{\rm divergent~polynomial
{}~+~finite~function} \Biggr) }
\eqno{(34)}
$$
\vskip .2in

\noindent (The $h_i$ represent logarithmic factors - those originally
present in the vertices of G times those from the coefficients of
the $T-T'$ polynomial.)

Lastly, the cutoff factors in front of the expressions above are
understood as follows.  The number of suppression factors coming from
all of the vertices in G,

$$
\delta(G)-\delta'(G)=\sum_{\rm vertices~in~G} (2m+2n-4)v^{(2n)}_{2m}
\eqno{(35)}
$$
\vskip .2in

\noindent must be greater than or equal to the number coming from
any portion of the graph G,

$$
\sum^a_{i=1}\biggl(\delta(\beta_i)-\delta'(\beta_i)
\biggr)=\sum_{{\rm vertices~in~all}~\beta_i} (2m+2n-4)v^{(2n)}_{2m}
\eqno{(36)}
$$
\vskip .2in

\noindent So

$$
\delta(G)-\delta'(G) \geq \sum^a_{i=1} \biggl(\delta(\beta_i)-
\delta'(\beta_i )\biggr) \eqno{(37)}
$$
\vskip .2in

\noindent and thus every term in ${\bar R}(G_k)$ in (34) has at
least the suppression $1/ \Lambda^2$ in front since $a\geq1$
for graphs containing at least one higher dimensional vertex.
All the finite functions in (34) are then divided by at
least $\Lambda^2$.

Overall each ${\bar R}(G_j)$ has a divergence that is local and
of degree $\delta'(G)=4-E$.  It also has finite parts suppressed
by factors of at least $1/ \Lambda^2$.  The original expression
for the renormalized graph, ${\bar R}'(G)$, is a sum of the terms
above in (34).  Hence ${\bar R}'(G)$ also has an overall primitive
divergence of degree $\delta'(G)=4-E$ which is local.  The remainder
is divided by at least $\Lambda^2$:

$$
R'(G)=(1-T'_G){\bar R}'(G) \leq {E^2\over\Lambda^2}
\hbox{\rm (~times~$\ln$s)}
\eqno{(38)}
$$
\vskip .2in

In actual practice none of this is necessary for we just have
to remove the divergences according to $\delta'$ power counting.
However, first they must be shown to be local in the external
momenta - thus the reason for the above analysis.

\vskip .4in
$\underline{(c)~General~Graphs}$
\vskip .4in

Lastly, we consider the most general Feynman diagrams - those that
contain nested divergences.  One proceeds recursively in subtracting
out all subdivergences in a general graph by replacing the lower
order subgraphs $U(\gamma)$ with ${\bar R}'(\gamma)$, their value
with all of their own subdivergences (according to $\delta '$) removed.

Recall that the ${\bar R}'(\gamma)$ above is written as a set of
graphs $\gamma_k$ (in equation (30) for example) with all of the
necessary counterterms to render their sub-integrations finite.
These graphs $\gamma$ may be subgraphs of some larger graph G, so to
continue the outlined procedure we rewrite the net subtraction to
remove the overall divergence in ${\bar R}'(\gamma)$ as:

$$
(-T'_{\gamma}){\bar R}'(\gamma)=\sum_k
(-T_{\gamma_k} +T_{\gamma_k}-T'_{\gamma_k}) \bar{R}(\gamma_k)
\eqno{(39)}
$$
\vskip .2in

\noindent The sum over $k$ symbolizes the different ways in which the
original graph $\gamma$ has been partitioned into new graphs, which
depends on the various ways we collect terms from the polynomials
$T-T'$ acting on $\gamma$'s subgraphs.  This can be done inductively
for more complicated graphs.  The explicit recipe for rewriting
the subtraction in (39) is presented in Appendix A, which holds
inductively in the recursive procedure to higher perturbative
order.  In section III(b) the rewriting (or adding and subtracting
procedure) was demonstrated on the simpler graphs.

By replacing the lower order subgraphs with their renormalized
counterparts we arrive at the recursion formula similar to the
one (27):

$$
{\bar R}'(G)=U(G)+\sum_{{ \{\gamma_1,\ldots,\gamma_n\}\in
P(G)},{\gamma_i\cap\gamma_j=\emptyset}} U({G\over \{\gamma_1,
\ldots,\gamma_n\}}) \prod^{n_a}_{i=1} \Biggl( \sum_{a_i}
(-T_{\gamma_{i,a_i}} +T_{\gamma_{i,a_i}}
-T'_{\gamma_{i,a_i}}){\bar R}(\gamma_{i,a_i}) \Biggr) \eqno{(40)}
$$
\vskip .2in

Equation (40) is the original Bogoliubov recursion formula, but
with $\delta'$ subtractions.

Next, the recursion relation in equation (40) may be iterated in order to
renormalize the graph and generate all sets of new graphs, as before.
The iteration must be handled as previously -- factoring the new
graphs into sets which have no subdivergences, so that the convergence
theorem may be applied.  That this is possible is due to the fact
that $T-T'$ acting on any primitive divergence is a polynomial with
terms divided by at least $\Lambda^2$.  The example given in figure 5
illustrates this procedure for a more complicated graph, which is
written out in detail in order to demonstrate how $\delta'$ subtractions
may be 'converted' into sets of graphs renormalized with subtractions
according to $\delta$.  Note that in the example the five loop
'cage' subdiagram possesses only a local divergence, so we do not
worry about subtractions on its own subgraphs.  In this example we
explicitly see how the original graph G
breaks into seven sets of integrals which may
only possess primitive divergences.

The equation above in (40) may be expressed as:

$$
{\bar R}'(G)=\sum_{G_k} \Biggl( U(G_k) + \sum_{{ \{\gamma_1,
\ldots,\gamma_n\}\in P(G_k)},{\gamma_i\cap\gamma_j=
\emptyset}}  U({G_k\over \{\gamma_1,\ldots,\gamma_n\}})
\prod^n_{i=1} ( -T_{\gamma_i} \bar{R}(\gamma_i) ) \Biggr)
\eqno{(41)}
$$
\vskip .2in

\noindent The graphs $G_k$ are denoted by all topological ways in
which the subgraphs in G are acted upon by the $T-T'$ operation, and
further by the various terms in each of their corresponding polynomial.
The many $T-T'$ that contribute to the definition of $G_k$ may be
disjoint or nested.  Correspondingly, each $G_k$ and its subtractions
may be viewed as a graph similar to G but with certain renormalized
subgraphs ${\bar R}(\Gamma)$ of G replaced with various (monomial)
higher dimensional vertices.

Since each of the terms in the sum over $G_k$ in (41) has no
subdivergences, it must have at most a polynomial divergence in
its loop integrations.  Counting powers of the cutoff tells us
that the net divergence in the graph $G_k$ is measured by
$\delta'=4-E$, E being the number of external legs, with a part
remaining that is divided by at least $\Lambda^2$.
Explicitly ${\bar R}'(G)$ in (41) takes on the form:

$$\eqalign{
{\bar R}'(G) & = {h_1(ln{\Lambda\over m})\over \Lambda^{\sum
(2m+2n-4)v^{(2n)}_{2m}}} \Biggl(\hbox{\rm
divergent~polynomial~of~degree~$\delta(G)$~+~finite~function~
of~dim.~$\delta(G)$} \Biggr)
\cr &
+{h_2(ln{\Lambda\over m})\over \Lambda^{\sum
(2m+2n-4)v^{(2n)}_{2m} -2}}
\Biggl(\hbox{\rm
divergent~polynomial~of~degree~$\delta(G)-2$~+~finite~function
{}~of~dim.~$\delta(G)-2$} \Biggr)
\cr &
+\ldots+{h_n(ln{\Lambda\over m})\over \Lambda^2} \Bigl(
\hbox{ \rm divergent~polynomial~of~order~4-E+2~+~finite~function~of
{}~dimension~(4-E+2)} \Bigr) }
\eqno{(42)}
$$

\noindent The subtraction that removes the overall divergence in
$\bar{R}'(G)$ according to $\delta'$ leaves:

$$
R'(G)=(1-T'_G){\bar R}'(G) \leq {E^2\over \Lambda^2}
(\hbox{\rm ~times~$\ln$s})
\eqno{(43)}
$$
\vskip .2in

\noindent Considering the form above in equation (41) or more
explicitly in (42), we may add and subtract the polynomial pieces
and write (43) also as:

$$
(-T'_G){\bar R}'(G)= \sum_k \Biggl( -T_{G_k} + T_{G_k} -
T'_{G_k} \Biggr) \bar{R}(G_k) \eqno{(44)}
$$
\vskip .2in

$\delta'$ naive power counting is justified inductively through the
use of Weinberg's theorem.  The form in (44) is exactly of the
type we had previously in (39), and in section III(b) it was shown explicity
on the simpler cases.  The fact that $T-T'$ acting on a primitive
divergence is a suppressed polynomial is responsible for the inductive
proof, since at each perturbative
order only new primitive divergences are
encountered.  These divergences are either renormalized or 'shrunk'
to a suppressed vertex.

We may simply reabsorb the divergences in the two- and four-point
Greens functions order by order in perturbation theory.  In actual
practice only the iterative procedure using $\delta'$ counting is
necessary.  The outlined roundabout procedure of adding and
subtracting polynomials to the original graph is there to establish
the perturbative effect of the irrelevant operators - that
they produce local divergences and contributions to the
renormalized Greens functions suppressed by at least $\Lambda^2$.

\vskip .3in
{\bf IV. Conclusions}
\vskip .3in

We have presented a method to understand the effects of irrelevant
operators in perturbative calculations.  By adding and
subtracting terms to a graph plus its two- and four-point
counterterms, thus rearranging the Feynman integrals, we have
effectively extended the BPH method to the case of power counting
with $\delta'$.  The renormalization of the general scalar theory
then follows by organizing the subtractions in the standard manner
of BPH.  The bounds on the graphs which contain the higher
dimensional vertices follow by keeping careful record of the
cutoff factors.

These operators ${1\over \Lambda^{2n}} O_i^{(2n)}~~(2n>2)$ produce
only local divergences in the two and four-point Greens functions
and corrections to a four dimensional field theory that are of order
$E^2/\Lambda^2$.  Our contribution here is by showing how this may
be understood through a natural generalization of the BPH approach.
Extensions to scalar theories in $d\neq 4$ dimensions, or to theories
not symmetric under $\Phi \rightarrow -\Phi$ is straightforward.

\vskip .3in
\noindent{\bf Acknowledgements}

It is a pleasure to thank Eric D'Hoker for helpful conversations
and guidance during the stages of this work.

\vfill
\break
\noindent{\bf Appendix~A}
\vskip .2in
Here we explicitly show how the general graphs with $\delta'$
subtractions explicitly break into renormalized (according to
$\delta$ power counting) ones, that:
\vskip .2in
$$
{\bar R}'(G) = \sum_{i} {\bar R} (G_i)
\eqno{(B1)}
$$
\vskip .2in
We start by using the explicit solution to the Bogoliubov recursion
formula in (1).  Denote a forest $F$ of a graph G to be a finite
set of 1PIR subgraphs $\gamma_i$ of G such that either:
\vskip .2in
$$
\gamma_i \subset \gamma_j, \quad \gamma_j \supset \gamma_i, \quad
or \quad \gamma_i \cap \gamma_j = \emptyset
\eqno{(B2)}
$$
\vskip .2in
\noindent Normal forests $F_{n}(G)$ of G are defined not to contain
the entire graph G and full forests do.  Then the explicit solution
to the Bogoliubov recursion formula is$^{(2)}$:
\vskip .2in
$$
{\bar R}'(G) = \sum_{U\in F_{n}(G)} ~\prod_{\gamma\in U}
\Bigl( -T_{\gamma}' \Bigr) U(G)
\eqno{(B3)}
$$
\noindent
and

$$
R'(G) = \sum_{U\in F(G)} ~\prod_{\gamma\in U}
\Bigl( -T_{\gamma}' \Bigr) U(G)
\eqno{(B4)}
$$
\vskip .2in
In (B4) $T_{\gamma}'$ is defined to be zero if $\delta'(\gamma)<0$,
and we consider all forests of G.  The subtractions
are to be performed inside to out in accord with the nested nature of
the elements in the forest.  Next we add and subtract
the additional subtractions so that:
\vskip .2in
$$
{\bar R}'(G) = \sum_{U\in F_{n}(G)} ~\prod_{\gamma\in U}
\Bigl( -T_{\gamma}'+ T_{\gamma} - T_{\gamma} \Bigr) U(G)
\eqno{(B5)}
$$
\vskip .2in
In order to prove the factorization we need to show that
(B5) splits into a sum over graphs $G_i$, defined below,
together with the full subtractions:
\vskip .2in
$$
\eqalign{
{\bar R}'(G) & = \sum_{i} \sum_{U\in F_{n}(G_i)} ~\prod_{\gamma\in U}
\Bigl( - T_{\gamma} \Bigr) U(G_i)
\cr &
= \sum_i {\bar R}(G_i)
}
\eqno{(B6)}
$$
\vskip .2in
We start by expanding the expression (B5) while respecting the order
of the nested differentiations.  This generates terms labeled by how
we may act $T-T'$ on different sets $\{ \gamma_1, \gamma_2,
\ldots, \gamma_n \}$ of 1PIR subgraphs of G.  The different sets are
conveniently labeled by the forests of G, so in fact the expansion
extends over all possible forests.

In order to continue we define $F_S (G)$ to be a forest of G not
including the elements $S=\{\gamma_1, \gamma_2, \ldots,
\gamma_n\}$.  Then the sum (B5) breaks into:
\vskip .2in
$$
{\bar R}'(G) = \sum_{S\in F(G)} ~\prod_{\beta\in S} \Bigl( T_{\beta} -
T_{\beta}' \Bigr) ~\sum_{U\in F_S (G)} ~\prod_{\gamma\in U}
\Bigl( - T_{\gamma} \Bigr) U(G)
\eqno{(B6)}
$$
\vskip .2in
\noindent Recall that all subtractions are to performed from the most
nested on out.  Let's switch the order of the subtractions to
follow this rule.  Define the maximal elements in $S$ to be
the set $S_m = \{ \gamma^{1}, \gamma^{2}, \ldots, \gamma^{m_s} \}$ such
that all elements of $S$ are contained in them and for all $i$ and
$j$ that $\gamma^{i} \cap \gamma^{j} = \emptyset$.  Then (B6) is:
\vskip .2in
$$
{\bar R}'(G) = \sum_{S\in F(G)} ~\sum_{U\in F(G/S_m)}
{}~\prod_{\gamma\in U} \Bigl( -T_{\gamma} \Bigr) ~
\Biggl( \prod_{\beta\in S} \Bigl( T_{\beta} - T_{\beta}' \Bigr)
\sum_{V\in F_S (S_m)} ~\prod_{\alpha\in V} \Bigl( -T_{\alpha} \Bigr)
\Biggr)
U(G)
\eqno{(B7)}
$$
\vskip .2in
The new graph $U(G_i)$, where $i$ denotes a forest $S$ of G and
individual terms in all of the subtractions $T_{\beta}-
T_{\beta}'$, is:
\vskip .2in
$$
U(G_i) =
\prod_{\beta\in S} \Bigl( T_{\beta} - T_{\beta}' \Bigr)
\sum_{V\in F_S (S_m)} \prod_{\alpha\in V} \Bigl( -T_{\alpha} \Bigr)
U(G)
\eqno{(B8)}
$$
\vskip .2in
\noindent The subtraction operators $T-T'$, by virtue of the
forest formula, must operate on only primitive divergences
since all subdivergences have been removed by subtractions $T$.
The 1PI elements of the maximal set $S_m$ have been shrunk to
points, and the forests of $G_i$ are exactly the forests of
$G/S_m$, $F(G_i)=F({G\over S_m})$.  Furthermore, the power
counting is not altered when multiple operations $T-T'$
are nested.  For two graphs $\alpha\subset\beta$, the effective
superficial degree of divergence for the
reduced graph $U(\beta_i)=U({\beta\over\alpha})
(T_{\alpha}-T_{\alpha}')\vert_{term~i}
\bar{R}(\alpha)$ satisfies $\delta(\beta) \geq \delta(\beta_i)$.  This
is a consequence of the fact that the $T-T'$ operation always results
in a polynomial whose terms are suppressed by hard powers of the cutoff.

As an example consider the simplest case, when $S_m=S$.  All elements
$\{ \gamma_1, \ldots, \gamma_n \}$ in S are maximal (i.e. $\gamma_i
\not \subset \gamma_j$), and expression (B8) splits into:
\vskip .2in
$$
\eqalign{
U(G_i) & = U({G\over S}) ~\prod_{\beta\in S} \Bigl(
(T_{\beta}-T_{\beta}') \sum_{V\in F(\beta)}
{}~\prod_{\alpha\in V} (-T_{\alpha})U(\beta) \Bigr)
\cr &
= U({G\over S}) ~\prod_{\beta\in S} (T_{\beta}-T_{\beta}')
\bar{R}(\beta)
}
\eqno{(B9)}
$$
\vskip .2in
\noindent All subdivergences in the 1PI parts of $S$ are eliminated, and
the parts themselves are replaced with higher-dimensional vertices.

Further examples are slightly more complicated, as some of the $T-T$
operations are nested.  Topologically, however, all subdivergences
are removed by the T operation in the sum in (B6).  The forest $S$ is
divided into disjoint maximal elements $\gamma^i$ which may be further
sub-divided into {\it their} maximal sub-elements.  Further sub-divisions
cease when the set is reached which is itself maximal.
Then the $T-T'$ operations on this maximal
set are polynomials divided by powers of
the cutoff.  Working backwards from
the most nested subgraphs, the $T-T'$ operations
always act on subgraphs which are fully subtracted, hence
lead to polynomials suppressed by at least $\Lambda^2$.

We have thus arrived at (B1).  The sum over forests of G and
individual terms in the $T-T'$ operations, denote
the possible new graphs $G_i$.

\vfill
\break
\noindent{\bf References}
\vskip .3in

\noindent (1)   N. Bogoliubov and O. Parasiuk, Acta Math v.97 (1957) 227;

     K. Hepp, Comm. Math. Physics v.2 (1966) 301;

     W. Zimmermann, Comm. Math. Physics v.15 (1969) 208

\noindent (2)   W. Zimmermann, "Brandeis Summer Lectures," (eds. S. Deser,
M. Grisaru, and H. Pendleton),

     MIT Press, 1970.

\noindent (3)   $\underline{\rm Renormalization}$, J. Collins (Cambridge
University Press, 1984);

     $\underline{\rm Quantum~Field~Theory}$, C. Itzykson and J. Zuber,
(McGraw-Hill Inc., 1980)

\noindent (4)   K.G. Wilson, J. Kogut, Physics Reports v.12, \#12 (1974) 75;

     K.G. Wilson, Phys. Rev. B4 (1970) 3174

\noindent (5)   J. Polchinski, Nucl.Phys. B231 (1984) 269

\noindent (6)   B. Warr, Annals of Physics v.183 (1988) 1

     B. Warr, Annals of Physics v.183 (1988) 59

\noindent (7)   S. Weinberg, Phys. Rev. v.118 (1960) 838
\vfill
\break

$$
\epsfxsize6.5in\vcenter{\epsfbox{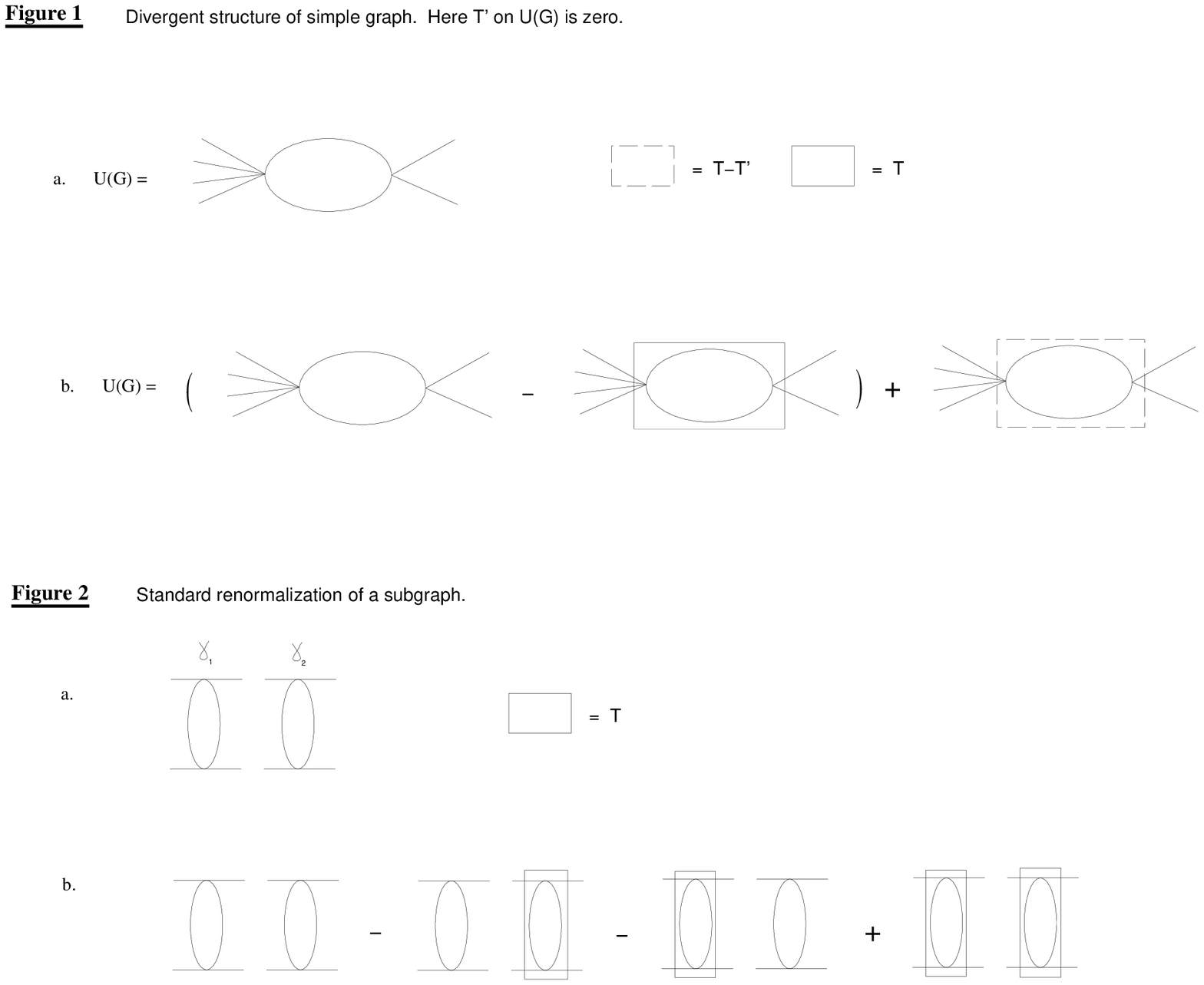}}
$$
\vfill\break
$$
\epsfxsize6.5in\vcenter{\epsfbox{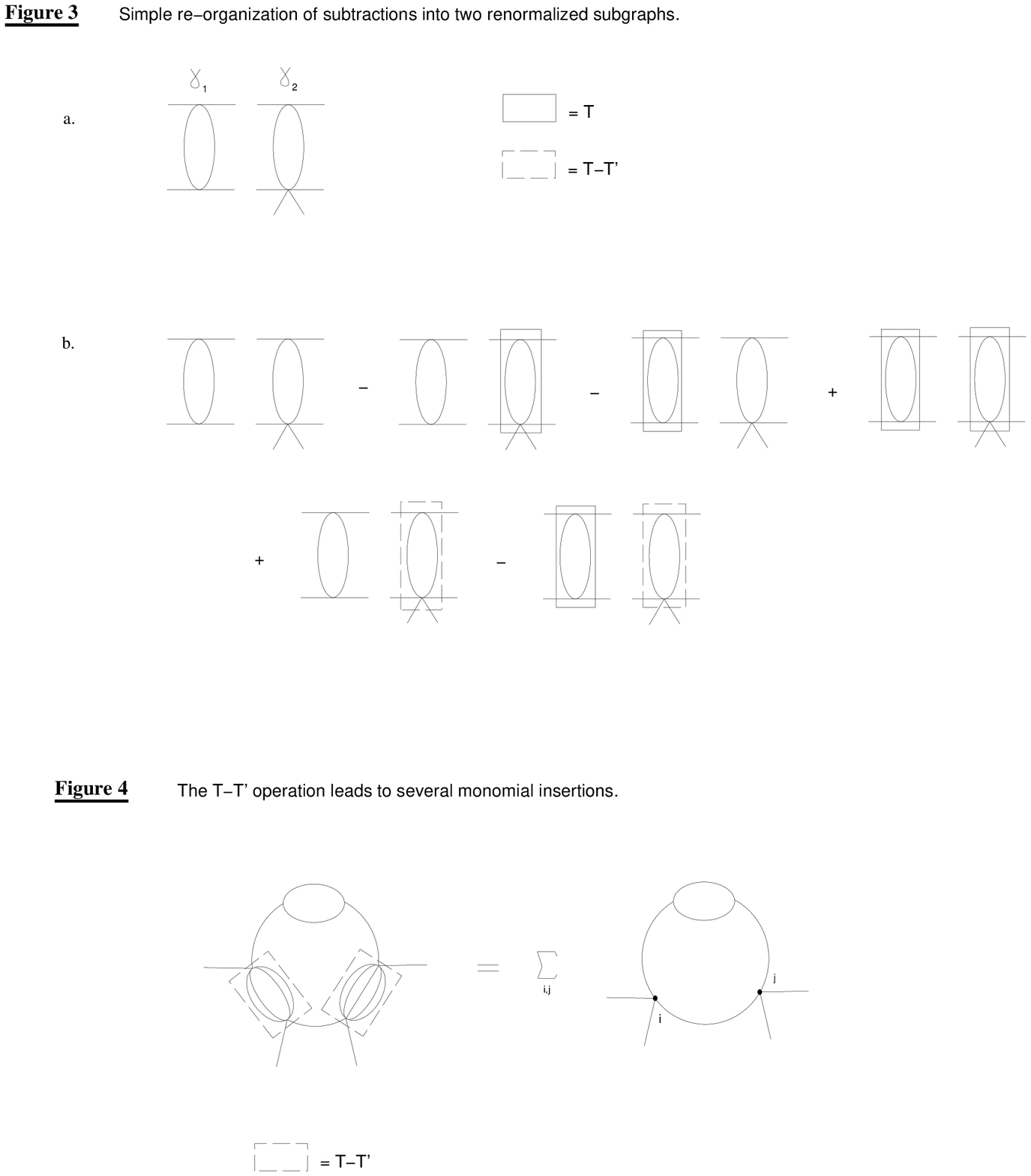}}
$$
\vfill\break
$$
\epsfxsize6.5in\vcenter{\epsfbox{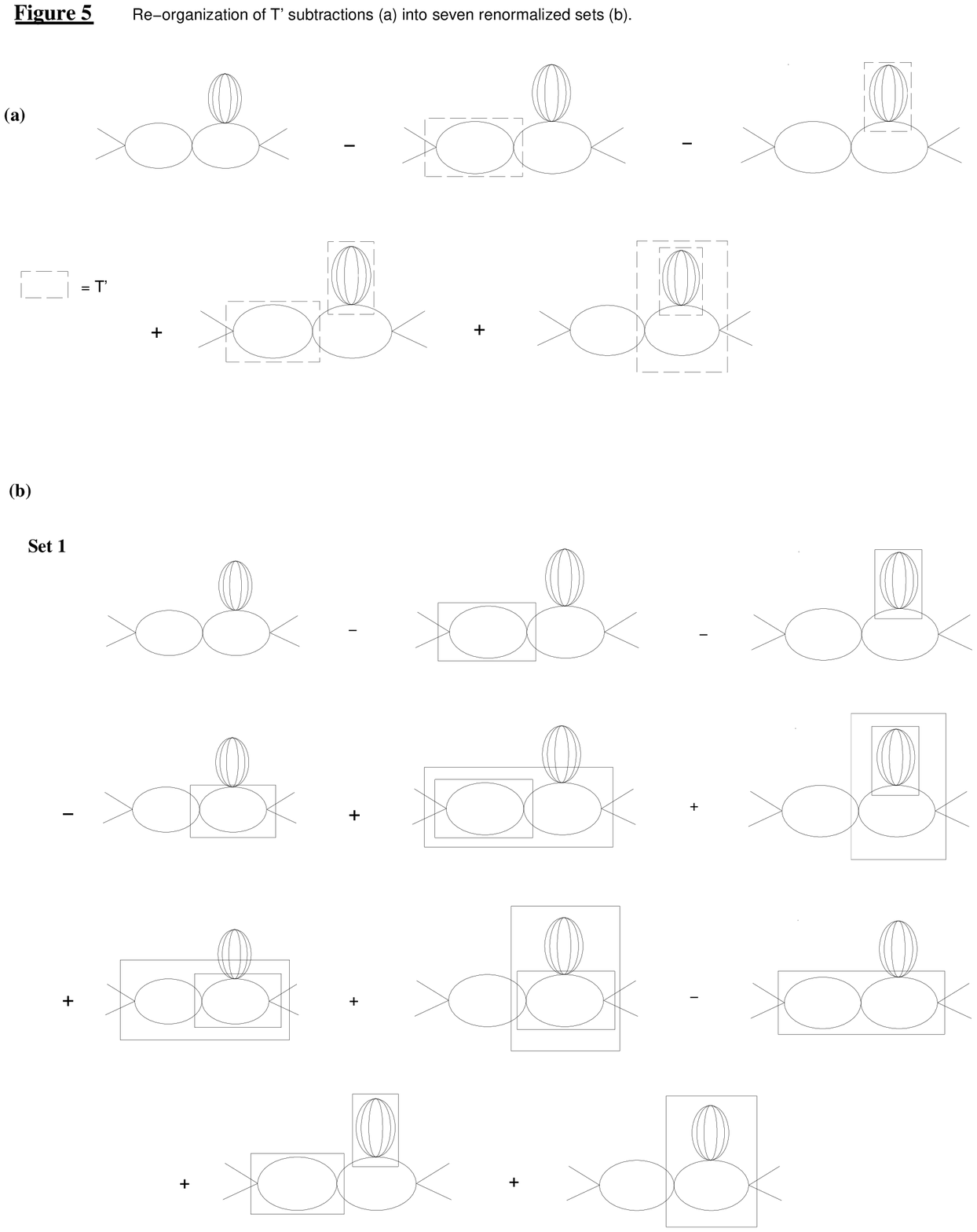}}
$$
\vfill\break
$$
\epsfxsize6.5in\vcenter{\epsfbox{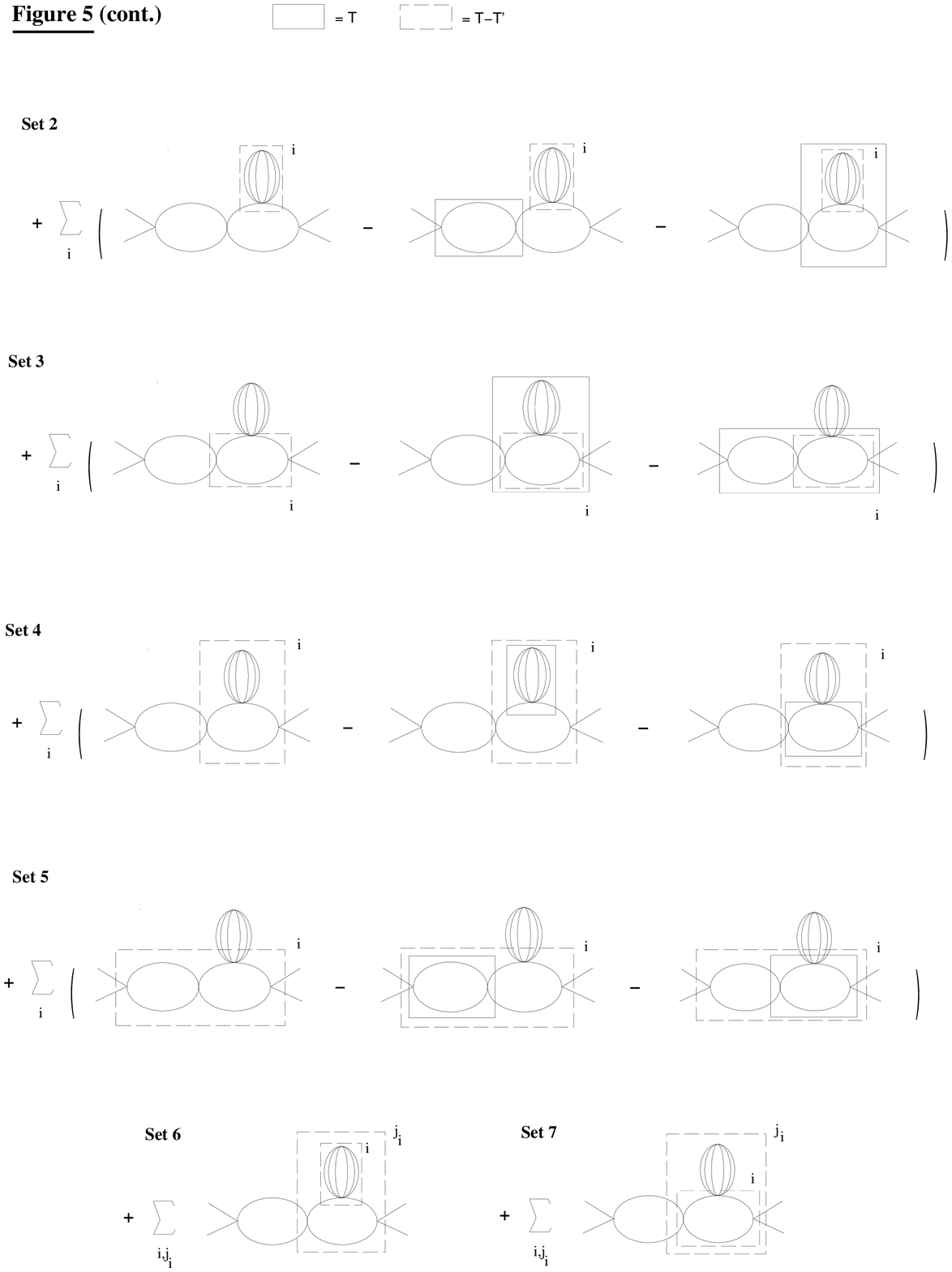}}
$$
\vfill\break
\end